\documentclass[a4paper,twoside]{article}
\newcommand{\affil}[1]{$^{\rm #1}$}
\usepackage[authoryear]{natbib}
\bibpunct{(}{)}{;}{a}{}{,}
\usepackage{graphicx}
\date{} 
\title{\large\bf\flushleft Neutron capture cross sections for the weak s process}
\author{\parbox{\textwidth}{\flushleft
\vspace{-0.5cm}
{\it M. Heil\affil{A,G}, A. Juseviciute\affil{B}, F.
K\"appeler\affil{B}, R.
Gallino\affil{C,D}, M. Pignatari\affil{E,F}, and E. Uberseder\affil{F}}\\
\vspace{0.4cm}
{\small \affil{A}\,Gesellschaft f\"ur Schwerionenforschung (GSI), D-64291 Darmstadt, Germany}\\
{\small \affil{B}\,Forschungszentrum Karlsruhe, Institut f\"ur Kernphysik, D-76021Karlsruhe, Germany}\\
{\small \affil{C}\,Dipartimento di Fisica Generale, Universit\`a di
Torino, via P. Giuria 1, 10125 (To), Italy}\\
{\small \affil{D}\,Center for Stellar and Planetary Astrophysics, Monash University, Victoria 3800, Australia}\\
{\small \affil{E}\,Astrophysics Group, School of
Physical and Geographical Sciences, Keele University, UK}\\
{\small \affil{F}\,University of Notre Dame, Notre Dame, USA}\\
{\small \affil{G}\,Email: m.heil@gsi.de}}}
\begin{document}
\twocolumn[
\begin{changemargin}{.8cm}{.5cm}
\begin{minipage}{.9\textwidth}
\vspace{-1cm}
\maketitle
%
%
\small{\bf Abstract:} In the past decades, a lot of progress has
been made for the understanding of the main s-process component
which takes place in TP-AGB stars. During this process about half of
the heavy elements, mainly between 90$\le$A$\le$209 are synthesized.
Improvements were made in stellar modeling as well as in measuring
relevant nuclear data for a better description of the main s
process. The weak s process, which contributes to the production of
lighter nuclei in the mass range 56$\le$A$\le$90 operates in massive
stars (M $\ge$8 M$_\odot$) and is much less understood. A better
characterization of the weak s-component would help disentangle the
various contributions to the element production in this region. For
this purpose, a series of neutron capture cross section measurements
on medium mass nuclei have been performed at the 3.7 MV Van de
Graaff accelerator at FZK using the activation method. Also, neutron
captures on abundant light elements with A$<$56 play an important
role for s-process nucleosynthesis, since they act as neutron
poisons and affect the stellar neutron balance. New results are
presented for the ($n, \gamma$) cross sections of $^{41}$K and
$^{45}$Sc, and revisions are reported for a number of cross sections
based on improved spectroscopic information.


\medskip
\medskip
\end{minipage}
\end{changemargin}
]
\small

\section{Introduction}
Since 1957, after the pioneering work by Burbidge, Burbidge, Fowler,
and Hoyle \cite{BBF57}, it is known that the elements heavier than
iron are mainly produced by two neutron capture processes, the
s(slow)- and the r(rapid)-process, both contributing about half of
the observed solar abundances between Fe and U. A third process, the
so called p(photodissociation)-process is responsible for the origin
of about 30 rare, proton-rich nuclei, but does not contribute
significantly to the synthesis of the elements in general ($<$1\%).

The main $\it{s}$ process produces predominantly nuclei with mass
numbers A$>$90 and is by far the most studied process. In recent
years, a detailed stellar model was developed \cite{SGB95} which
suggests that the main component of the $\it{s}$ process occurs in
the He-rich intershell of thermally pulsing AGB stars and the
calculated s-abundances are in excellent agreement with observations
\cite{AKW99b}.

The weak $\it{s}$ component, on the other hand, is much less
understood. It is responsible for the production of nuclei between
iron and yttrium (56$\le$A$\le$90). Current stellar models suggest
that it takes place during convective core-He burning in massive
stars (M $\ge$ 8M$_\odot$), where temperatures of
(2.2-3.5)$\times$10$^{8}$ K at the center are reached near He
exhaustion and neutrons are liberated by the activation of the
$^{22}$Ne($\alpha, n$)$^{25}$Mg reaction. Since the neutron exposure
is small, the $\it{s}$-process flow can not overcome the bottleneck
at the closed neutron shell N=50. Most of the material is
reprocessed by the following burning stages and only a small part
survives in the outer layers of the previous convective core He
burning and is ejected during the supernova explosion. A second
neutron exposure occurs during convective carbon shell burning of
massive stars \cite{RGB93, RHH02, TEM07}. There, neutrons are
produced mainly by burning leftover $^{22}$Ne. Since $^{22}$Ne
derives from initial CNO nuclides, first converted to $^{14}$N by H
burning and then by the chain
$^{14}$N($\alpha$,$\gamma$)$^{18}$F($\beta^+$$\nu$)$^{18}$O($\alpha$,$\gamma$)$^{22}$Ne
occurring in the first phase of core He burning, the weak s-yields
decrease with metallicity. The high temperatures during carbon
burning of $\it{T}\sim$1$\times$10$^9$ K cause high neutron
densities up to $\sim$10$^{12}$ cm$^{-3}$.

The nucleosynthesis yields of the weak component in massive stars
are also important for the $\it{r}$ process supernova scenarios,
since they determine the composition of a star before the supernova
explosion. Since the $\it{s}$-process abundances can be determined
reliably on the basis of experimental ($n,\gamma$) cross sections,
the $\it{r}$ abundances are commonly inferred by the
$\it{r}$-residual method, that is by subtracting the $\it{s}$
abundances from solar values, N$_r$=N$_{\odot}$$-$N$_s$
\cite{AnG89}. The $\it{r}$ abundances obtained in this way are then
used for testing $\it{r}$-process models. This is of special
interest, since recent observations of ultra-metal-poor halo stars
\cite{SCL03} suggest a second, so-called weak $\it{r}$ process,
which contributes to the element production below barium.

In addition, current $\it{s}$-process models cannot explain the high
observed abundances of the typical s elements Sr, Y, and Zr in halo
stars with low metallicity \cite{TGA04}. Significant contributions
by the main $\it{s}$ process to the interstellar medium are to be
expected at [Fe/H]$>$$-$1.5 only, because of the long lifetimes of
low mass AGB stars. Therefore, a new primary neutron capture
process, the lighter element primary process (LEPP) was suggested by
Travaglio et al. \cite{TGA04}. Another scenario pointing to a
s-process origin of LEPP is that this extra component is produced by
fast rotating metal poor massive stars. The rotation mixes primary
$^{14}$N in the He core where it is converted in primary $^{22}$Ne,
causing a strong production of neutrons before the He
exhaustion and in the following C shell \cite{PGM08}.\\
Recently, a totally new process, the $\nu$$\it{p}$ process, was
introduced \cite{FML06}, where nucleosynthesis calculations explored
the so far neglected effect of neutrino interactions and found that
it is possible to produce neutrons via antineutrino captures on
protons in the innermost proton-rich ejecta of core-collapse
supernovae. Neutron densities of 10$^{14}$ - 10$^{15}$ cm$^{-3}$
could be obtained in this way for several seconds, when the
temperatures are in the 1-3 GK range.

It is clear from the above discussion that several processes
contribute to the nucleosynthesis of medium-heavy nuclei from Fe to
Ba. In order to disentangle the various processes and to identify
possible astrophysical sites, the individual contributions have to
be separated. A first step could be the identification of the
abundance contributions from the weak $\it{s}$ process, since the
most important nuclear data input, the neutron capture cross
sections, are accessible by laboratory experiments. Therefore, new
neutron capture cross section measurements on medium mass nuclei are
demanded.

The cross sections of light elements are also important in this
respect because they affect the neutron balance inside stars during
the $\it{s}$ process. Although their cross sections are small, these
elements are much more abundant than those in the mass region above
Fe. Therefore, light elements constitute potential neutron poisons
and may consume neutrons, which are then not available for
$\it{s}$-process nucleosynthesis. Especially important in this
respect are neutron captures on $^{12}$C, $^{16}$O and on the neon
and magnesium isotopes, but also other light isotopes up to iron
contribute as well. In many of these cases the neutron capture cross
sections are not known with sufficient accuracy since they are small
and difficult to measure.

Neutron capture cross sections of light isotopes play also an
important role for analyses of presolar grains, which can provide
stringent constraints on $\it{s}$-process models \cite{Zin98}.
Because these grains are only a few $\mu$m in size and because the
abundances of heavy elements are rather low, their isotopic
composition is difficult to determine for individual grains. Lighter
elements are more abundant and, therefore, easier to detect.

In this contribution we report on neutron capture cross section
measurements of $^{41}$K and $^{45}$Sc. The experimental method and
the results are described in chapter 2. The measurements are part of
a series of neutron capture studies on light and medium heavy nuclei
relevant for the weak $\it{s}$ process in massive stars, which will
be summarized in chapter 3. Section 4 concludes with some remarks on
the astrophysical implications of these results.

\section{Activation method}

The most important input for stellar models of the $\it{s}$ process are
Maxwellian averaged neutron capture cross sections (MACS) and
$\beta$-decay rates, but also stellar enhancement factors (SEF) have
to be known.

The weak $\it{s}$ process contributes substantially to the production of
elements in the mass range 56$\leq$A$\leq$90. In this mass region the
($n, \gamma$) cross sections show large uncertainties and need significant
improvement for a reliable description of the abundance contributions
from massive stars. This is especially important since the local
approximation ($\langle \sigma \rangle$N = const) is not valid
during the weak $\it{s}$ process. Therefore, any change in the cross
section of a light isotope, e.g. $^{62}$Ni \cite{NPA05}, can affect
the abundances of all the heavier isotopes up to zirconium and maybe
even higher up. This underlines that neutron capture cross sections
in the mass range 50$\leq$A$\leq$90 have to be measured with
significantly higher accuracy.

\begin{figure}[h]
\begin{center}
\includegraphics[scale=0.48, angle=0]{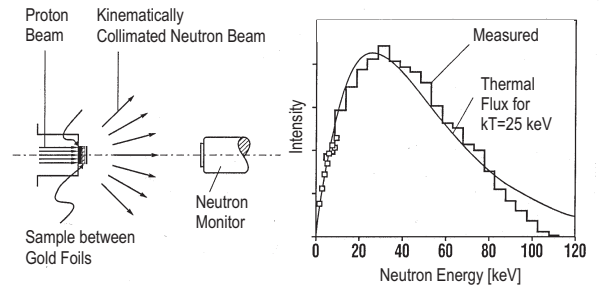}
\caption{The experimental setup for activation measurements (left)
and a comparison of the produced neutron spectrum and a thermal
neutron spectrum at $kT$=25 keV (right). \label{fig1}}
\end{center}
\end{figure}

Therefore, a measuring campaign was launched at Forschungszentrum
Karlsruhe with the aim to improve the neutron capture cross sections
of light and medium mass nuclei. A reliable and accurate approach to
determine Maxwellian averaged cross sections at $kT$=25 keV is the
activation method \cite{BeK80}, where the $^7$Li($p, n$)$^7$Be reaction
is used to produce a quasi-stellar neutron spectrum as sketched in
Fig. \ref{fig1}. After irradiation in that spectrum the induced
sample activity is counted in a low background environment with high
resolution Ge detectors. The proton beam with an energy of $E_p$=1912
keV and typical intensities of 100 $\mu$A was delivered by the
Karlsruhe 3.7 MV Van de Graaff accelerator. The neutron production
target consists of a metallic Li layer, which is evaporated onto a
water cooled copper backing. The sample is placed inside the resulting
neutron cone, which has an opening angle of 120 degrees in the
direction of the proton beam. The neutron flux is monitored
throughout the irradiations by means of a $^{6}$Li-glass detector,
positioned at a distance of 1 m from the target. After the
irradiation the total number of activated nuclei A is given by
\begin{equation}
A = \Phi \cdot N \cdot \sigma \cdot f_{b},
\end{equation}
\noindent where $\Phi$ is the time integrated neutron flux, $N$ the
number of sample atoms per cm$^2$, and $\sigma$ the spectrum
averaged neutron capture cross section. In order to determine the
neutron flux, the sample is sandwiched between gold foils. Since the
gold cross section is well known, the total number of neutrons can
be obtained by measuring the 412 keV line from the decay of
$^{198}$Au with HPGe detectors. The factor $f_{b}$ accounts for the
variation of the neutron flux and for the decay during activation.
The cross section is then calculated from the number of counts
in a characteristic $\gamma$-ray line
\begin{equation}
C_{\gamma}=A\cdot K_{\gamma} \cdot \varepsilon _{\gamma} \cdot
I_{\gamma} \cdot (1-exp(-\lambda t_{m}))\cdot exp(-\lambda t_{w}),
\end{equation}
\noindent where $K_{\gamma}$ is a correction factor for $\gamma$-ray
self-absorption, $\varepsilon _{\gamma}$ the efficiency of the
Ge-detector, $I_{\gamma}$ the line intensity, $t_{w}$ the waiting
time between irradiation and counting, and $t_{m}$ the duration of
the activity measurement.

\begin{table*}[h]
\caption{Summary of irradiation parameters.}
\label{samples}
\begin{tabular}{@{}cccccc}
\hline
 Activation & Sample      & Mass   & Diameter & Irradiation time & Integrated neutron\\
            & composition & (mg)   & (mm)     & (min)             & flux  (10$^{14}$)\\
\hline
1           & KBr         & 98.70  &  6       & 156              & 0.2681        \\
2           & KBr         & 733.27 & 12       & 1118             & 0.7069        \\
3           & Sc metal    & 2.43   &  6       & 2829             & 5.019         \\
4           & Sc metal    & 6.37   & 10       & 4148             & 3.352         \\
5           & Sc metal    & 9.57   & 12       & 2829             & 1.855         \\

\hline
\end{tabular}\\[2pt]
\end{table*}

In this way, we have measured the MACSs of several nuclei. As an
example, we give here details of the measurements on $^{41}$K, and
$^{45}$Sc at $\it{kT}$=25 keV. A brief summary of all measurements
follows in Section 3.

By variation of sample dimensions and other basic experimental
parameters in repeated activations it was possible to verify the
data analysis procedures and to check the reliability of
the evaluated uncertainties of the measurements. The information
on the relevant parameters of the different activation runs are
given in Table \ref{samples}. The decay properties of the
product nuclei, which are essential for determining the induced
activities are summarized in Table \ref{decay_properties}. A
detailed description of the procedures used in the measurements
and in data analysis can be found in \cite{HKU08a} and references
therein.

\begin{table*}[h]
\caption{Decay properties of the product nuclei.}
\label{decay_properties}
\begin{tabular}{@{}ccccc}
\hline
 Product   & Half-life       & $\gamma$-ray energy  & Intensity per decay   & Reference\\
 nucleus   & (m)             & (keV)                & (\%)                  &          \\
\hline
$^{42}$K   & 741.6$\pm$0.72  & 1524.6               & 18.08$\pm$0.09        & \cite{SiC01} \\
$^{46}$Sc  & 120658$\pm$58   & 889.28               & 99.984$\pm$0.001      & \cite{Wu00}  \\
           &                 & 1120.55              & 99.987$\pm$0.001      &              \\
$^{198}$Au & 3881.1$\pm$0.29 & 411.8                & 95.6$\pm$0.12         & \cite{Chu02} \\
\hline
\end{tabular}\\[2pt]
\end{table*}

\begin{table*}[h]
\caption{Measured cross sections.}
\label{results}
\begin{tabular}{@{}ccccc}
\hline
  Activation & $\gamma$-ray energy & Self-absorption & Cross section & Mean value \\
             & (keV)               & factor          & (mbarn)          & (mbarn)       \\
\hline
\multicolumn{5}{c}{$^{41}$K($n, \gamma$)$^{42}$K}\\
  1          & 1524.6              & 0.99            & 20.4$\pm$1.6$\pm$0.7 & \\
  2          &                     & 0.98            & 20.2$\pm$0.5$\pm$0.6 & 20.3$\pm$0.8(stat)$\pm$0.7(sys) \\
             &                     &                 &                      &                                 \\
\multicolumn{5}{c}{$^{45}$Sc($n, \gamma$)$^{46}$Sc}\\
  3          & 889.28              & 1.00            & 59.3$\pm$0.9$\pm$1.8 & \\
             & 1120.55             & 1.00            & 61.7$\pm$1.0$\pm$1.9 & \\
  4          & 889.28              & 1.00            & 59.7$\pm$0.7$\pm$1.8 & \\
             & 1120.55             & 1.00            & 61.0$\pm$0.7$\pm$1.8 & \\
  5          & 889.28              & 1.00            & 60.3$\pm$0.7$\pm$1.8 & \\
             & 1120.55             & 1.00            & 60.0$\pm$0.8$\pm$1.8 & 60.3$\pm$0.4(stat)$\pm$1.8(sys) \\
\hline
\end{tabular}\\[2pt]
\end{table*}

The experimental results and uncertainties are summarized in Tables
\ref{results} and \ref{uncertainties}. In spite of the variation of the
experimental parameters, the results are all consistent within the
estimated uncertainties, thus confirming the reliability of the
experimental method. These variations included different sample sizes and
masses to verify the corrections for finite size and self shielding
effects as well as different irradiation times to control
uncertainties due to the half-lives of the respective product
nuclei. Significant contributions to the overall uncertainty
originate from the gold reference cross section, the efficiency of
the HPGe detectors, and the time integrated neutron flux, whereas
the uncertainties from the $\gamma$-decay intensities are small.

\begin{table}[h]
\caption{Compilation of systematic uncertainties.
\label{uncertainties}}
\begin{tabular} {lccc}
\hline
 Source of Uncertainty & \multicolumn{3}{c}{Uncertainty (\%)}  \\
                                  & Au      & $^{41}$K   & $^{45}$Sc \\
\hline
Gold cross section                & 1.5     &  -         & -       \\
Number of nuclei                  & 0.6     &  0.1       & 0.4     \\
Time factors $f_b$, and $t _{1/2}$& $<$0.1  & $<$0.1     & $<$0.1  \\
Self-absorption                   & $<$1.0  & $<$1.0     & $<$1.0  \\
Detector efficiency               &  1.5    & 1.5        & 1.5     \\
$\gamma$-ray intensity per decay  &  0.13   & 0.5        & 0.001   \\
Neutron flux                      & -       & 2.6        & 2.4     \\
                                  &         &            &         \\
Total uncertainty                 & -       & 3.2        & 3.0     \\
\hline
\end{tabular}
\end{table}

The results in Table \ref{results} represent cross section values
averaged over the quasi-stellar neutron spectrum used for the
irradiations. This spectrum is close but not identical to a
Maxwellian distribution. In the experimental spectrum we have e.g. a
cut-off at an energy of 106 keV. In order to derive MACSs which are
defined as

\begin{equation}
<\sigma>_{kT}=\frac{<\sigma v>}{v_T}=
 \frac{2}{\sqrt{\pi}}
\frac{\int_0^{\infty} \sigma(E_n)\cdot E_n \cdot {\rm
exp}(-E_n/kT)\cdot dE_n} {\int_0^{\infty} E_n \cdot {\rm
exp}(-E_n/kT)\cdot dE_n}
\end{equation}

we used the evaluated energy-dependent cross sections,
$\sigma(E_n)$, from data libraries, which were normalized to
reproduce our experimental results. In the above formula $E_{n}$
denotes the neutron energy, $k$ the Boltzmann factor, and $T$ the
temperature. The MACSs obtained for various thermal energies are
listed in Tables \ref{41K} and \ref{45Sc} for the respective
libraries as well as for the temperature trend of the compilation by
Bao et al. (2000).

\begin{table}[h]
\caption{MACSs for the $^{41}$K($n, \gamma$)$^{42}$K reaction
calculated with the normalized neutron capture cross sections of
various databases compared with the temperature trend of Bao et al.
(2000).} \label{41K}
\begin{tabular}{@{}cccc}
\hline
 kT     & JEFF 3.1$^{a}$ & JENDL 3.3$^{b}$ & Bao et al. (2000) \\
(keV)   & (mbarn)         & (mbarn)         & (mbarn)                       \\
\hline
5       & 91.2            &  91.2           & 75.9$^{+16.3}_{-5.7}$ \\
10      & 50.7            &  50.7           & 42.5$^{+8.8}_{-3.2}$  \\
15      & 35.3            &  35.3           & 30.8$^{+5.1}_{-2.3}$  \\
20      & 26.7            &  26.6           & 24.8$^{+2.7}_{-1.9}$  \\
25      & 21.3            &  21.3           & 21.3$^{+1.6}_{-1.6}$  \\
30      & 17.8            &  17.8           & 19.0$^{+1.4}_{-1.9}$  \\
40      & 13.8            &  15.4           & 16.1$^{+1.2}_{-2.6}$  \\
50      & 11.7            &  11.7           & 14.4$^{+1.1}_{-2.9}$  \\
60      & 10.4            &  10.4           & 13.2$^{+1.0}_{-3.0}$  \\
80      & 9.1             &  9.1            & 11.6$^{+0.9}_{-2.6}$  \\
100     & 8.4             &  8.4            & 10.4$^{+0.8}_{-2.1}$  \\
\hline
\end{tabular}\\
$^{a}$ JEFF/3.1 (www.nea.fr/html/dbdata/JEFF/)\\
$^{b}$ JENDL/3.3 (wwwndc.tokai-sc.jaea.go.jp/jendl/)
\medskip\\
\end{table}

Since it is not obvious, which trend with $kT$ is to be preferred,
and since it is beyond the scope of this paper to trace the origin
of the differences between various evaluations, the recommended
values, which have been used for the $s$-process calculations
discussed in the following section, are the ones obtained using the
temperature trend of Bao et al. (2000). This choice was motivated by
the fact that these data include the most recent time-of-flight
(TOF) results. The uncertainties of the extrapolated MACS were
estimated by comparison with the upper and lower bounds obtained by
using the evaluated cross sections from the data libraries. The
uncertainties of the recommended values are, therefore, composed of
the experimental uncertainties originating from the measured data
and of the contributions defined by the differences with respect to
the values derived from different databases.

Since the MACS in Tables \ref{41K} and \ref{45Sc} have already been
normalized to the present cross section results, it is interesting
to note the normalization factors as well. For $^{41}$K these are
0.57, 0.57, and 0.86 for JEFF/3.1, JENDL/3.3, and Bao et al. (2000),
and for $^{45}$Sc 0.88, 0.85, 0.88, and 0.80 for JEFF/3.1,
ENDF/B-6.8, JENDL/3.3, and Bao et al. (2000), respectively.

The present results are consistently smaller compared to previous
data, which both have been obtained in TOF experiments. For the MACS
of $^{41}$K the only other experiment reports a 12\% larger value
\cite{Mac84b}, whereas the present and previous measurement claim
uncertainties of only 7.5\% and 3.2\%. In case of $^{45}$Sc there is
also only one other experimental value \cite{KAM77}, which is 25\%
larger, far outside the quoted uncertainties of 7.2\% and the
present 3.0\%.

\begin{table*}[h]
\caption{MACSs for the $^{45}$Sc($n, \gamma$)$^{45}$Sc reaction
calculated with the normalized neutron capture cross sections of
various databases compared with the ones obtained using the
temperature trend of Bao et al. (2000). \label{45Sc} }
\begin{tabular}{@{}cccccc}
\hline
 kT    & JEFF/3.1$^a$ & ENDF/B-6.8$^b$ & JENDL/3.3$^c$ &  Bao et al. (2000) \\
(keV)  & (mbarn)      & (mbarn)        & (mbarn)       &  (mbarn)           \\
\hline
5      & 163          & 161            & 163           &  181$^{+7}_{-21}$     \\
10     & 115          & 115            & 115           &  123$^{+5}_{-9}$      \\
15     & 89.7         & 88.8           & 89.7          &  92.9$^{+3.4}_{-5.3}$ \\
20     & 74.6         & 73.4           & 74.7          &  76.1$^{+2.8}_{-3.9}$ \\
25     & 64.3         & 63.2           & 64.7          &  64.1$^{+2.3}_{-2.3}$ \\
30     & 56.8         & 56.1           & 57.5          &  55.3$^{+3.0}_{-2.0}$ \\
40     & 46.4         & 46.9           & 47.7          &  43.3$^{+4.7}_{-1.6}$ \\
50     & 39.5         & 41.1           & 41.2          &  35.3$^{+6.0}_{-1.3}$ \\
60     & 34.4         & 36.8           & 36.6          &  30.4$^{+6.5}_{-1.1}$ \\
80     & 27.7         & 30.6           & 30.3          &  24.8$^{+5.9}_{-0.9}$ \\
100    & 23.4         & 26.0           & 26.2          &  21.6$^{+4.7}_{-0.8}$ \\
\hline
\end{tabular}\\
\medskip\\
$^{a}$ JEFF/3.1 (www.nea.fr/html/dbdata/JEFF/)\\
$^{b}$ ENDF/B-VI.8 (www.nndc.bnl.gov/)\\
$^{c}$ JENDL/3.3 (wwwndc.tokai-sc.jaea.go.jp/jendl/)
\end{table*}

\section{Neutron capture measurements}

With the activation technique described before, we have determined
a number of MACSs in the mass range of the weak $s$ process
between Fe and Zr. Apart from the measurements on $^{41}$K and
$^{45}$Sc presented here, the results for the other isotopes in
our study were published elsewhere. In order to provide an
overview all recent measurements are summarized in Table \ref{table1}
and compared with the previously recommended values \cite{BBK00}.

\begin{table*} [h]
\begin{center}
\caption{Recent MACS at $kT=30$ keV (in mbarn) compared with
previously recommended values. \label{table1}}
\begin{tabular}{cccc}
\hline Target isotope & Bao et al. (2000) & Present activations  & Reference \\
\hline
$^{19}$F  & 5.8$\pm$1.2  & 3.2$\pm$0.1          & \cite{UHK07}  \\
$^{41}$K  & 22.0$\pm$0.7 & 19.0$^{+1.4}_{-1.9}$ & This Work     \\
$^{45}$Sc & 69$\pm$5     & 55.3$^{+3.0}_{-2.0}$ & This Work     \\
$^{58}$Fe & 12.1$\pm$1.3 & 13.5$^{+0.6}_{-0.8}$ & \cite{HKU08a} \\
$^{59}$Co & 38$\pm$4     & 39.6$^{+2.7}_{-2.5}$ & \cite{HKU08a} \\
$^{64}$Ni & 8.7$\pm$0.9  & 8.0$^{+0.5}_{-0.8}$  & \cite{HKU08a} \\
$^{63}$Cu & 94$\pm$10    & 56.0$^{+2.2}_{-5.2}$ & \cite{HKU08a} \\
$^{65}$Cu & 41$\pm$5     & 30.0$^{+1.3}_{-1.9}$ & \cite{HKU08a} \\
$^{79}$Br & 627$\pm$42   & 613$\pm$59           & \cite{HKU08b} \\
$^{81}$Br & 313$\pm$16   & 235$\pm$9            & \cite{HKU08b} \\
$^{85}$Rb & 240$\pm$9    & 221$^{+13}_{-6}$     & \cite{HKU08b} \\
$^{87}$Rb & 15.5$\pm$1.5 & 15.8$^{+0.7}_{-0.9}$ & \cite{HKU08b} \\
\hline
\end{tabular}
\end{center}
\end{table*}

For some isotopes the measured cross sections differ significantly
from previous recommendations. It is conspicuous that many new
results are systematically smaller than the recommended cross
sections from Bao et al. (2000), which are often based on time-of-flight
(TOF) measurements performed with C$_{6}$D$_{6}$ detectors in the
1970ies and early 1980ies \cite{KAM77, Mac84b, MAM78b}. In fact, this trend
is confirmed by a general comparison between MACSs obtained with the
activation method and the TOF method performed with C$_{6}$D$_{6}$
detectors, which reveal large discrepancies on average. The cross
sections from activation measurements are consistently lower, often
in complete disagreement within the quoted uncertainties.

\begin{figure}[h]
\begin{center}
\includegraphics[scale=0.3, angle=0]{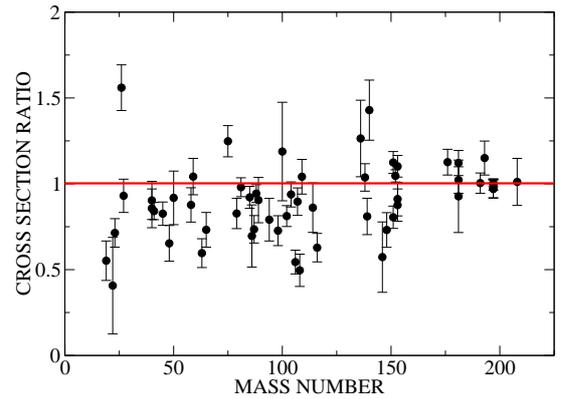}
\caption{Comparison of MACSs at $\it{kT}$=30 keV obtained by
activation and by the TOF method. Note that most ratios are
smaller than unity. \label{fig2}}
\end{center}
\end{figure}

This is illustrated in Fig. \ref{fig2} by the comparison
between MACSs at $\it{kT}$=30 keV obtained with the TOF and the
activation method. A possible explanation could be that the
background due to sample-scattered neutrons was systematically
underestimated in older TOF experiments. Neutrons scattered in
the sample and captured in the detector and/or in surrounding
materials produce background events, which are difficult to
distinguish from true capture events. This background can be as high
as 50\% for light and medium heavy nuclei, where the
scattering/capture ratios are large. The correspondingly large and
uncertain corrections tend to give rise to large systematic errors.

It should be also noted that some activation measurements performed
in the past are inconsistent with newer results. The neutron capture
cross section of $^{81}$Br can serve as an example. The MACS at
$\it{kT}$=30 keV of the activation measurement in 1986 of 317$\pm$16
mbarn \cite{WBK86b} is in contradiction to the new result of Heil et
al. \cite{HKU08b} of 235$\pm$9 mbarn. However, in contrast to the
previous result, which was measured in 1986 by a single activation,
several repeated activations were performed in Ref. \cite{HKU08b} by
variation of sample dimensions and other experimental parameters as
described before.

Since activation measurements have to rely on sometimes rather uncertain
decay properties of the respective product nuclei, any improvement of
these basic parameters such as decay intensities or half-lives are
important and can be used for the revision of older activation data.
A systematic search for improved decay information resulted in the
set of updated MACSs, which are listed in Table \ref{updated_MACS}
together with the corresponding correction factors and references to
the new decay data.

\begin{table*} [h]
\begin{center}
\caption{MACS at kT=30 keV (in mbarn). \label{updated_MACS}}
\begin{tabular}{ccccc}
\hline
Target isotope & Bao et al. (2000) & Correction factor & New MACS  & Ref. to new decay data\\
\hline
$^{46}$Ca   & 5.7$\pm$0.5$^a$  & 1.085   & 6.2$\pm$0.7 & \cite{Bur95}  \\
            & 4.9$\pm$0.6$^b$  & 1.042   & 5.1$\pm$0.5 &               \\
$^{124}$Xe  & 644$\pm$83       & 0.958   & 617$\pm$79  & \cite{Kat99}  \\
$^{155}$Eu  & 1320$\pm$84      & 0.883   & 1166$\pm$73 & \cite{Rei94}  \\
$^{152}$Gd  & 1030$\pm$65      & 0.930   & 958$\pm$47  & \cite{Hel98} \\
$^{158}$Gd  & 221$\pm$25       & 0.816   & 178$\pm$42  & \cite{Hel94} \\
$^{160}$Gd  & 144$\pm$14       & 1.618   & 230$\pm$16  & \cite{ReH00} \\
\hline
\end{tabular}
\medskip\\
$^a$ Correction for value in K\"appeler et al. (1985) \cite{KWM85}.\\
$^b$ Correction for value in Mohr et al. (1999) \cite{MSB99}.\\
\end{center}
\end{table*}

\section{Astrophysical implications}

\begin{figure}[h]
\begin{center}
\includegraphics[scale=0.3, angle=270]{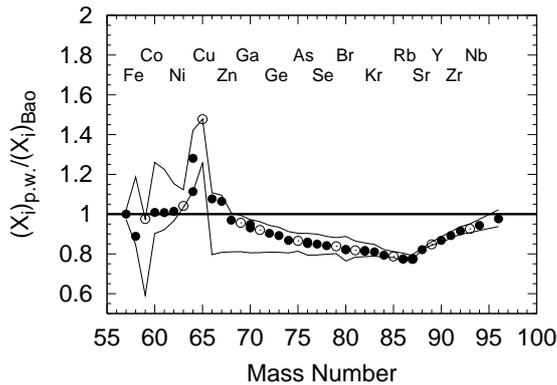}
\caption{Nucleosynthesis yields between Fe and Nb illustrating the
final s-process yields after shell C burning for a 25 M$_{\odot}$
star with [Fe/H] = 0. To illustrate the combined effect of all new
cross sections the yield is plotted relative to the standard case
using the cross sections of Bao et al. (2000) (p.w. stands for
present work, even and odd Z elements are distinguished by black and
open symbols, respectively). The thin lines correspond to the upper
and lower limits of the cross sections in Tables VI and VII and
demonstrate the uncertainties stemming from the extrapolation of the
measured cross sections to higher and lower energies. \label{fig03}}
\end{center}
\end{figure}

The MACS obtained in a series of neutron capture measurements with
the activation method have been used to explore their impact on
stellar model calculations for the weak $s$ process in a 25
M$_{\odot}$ star, which were performed with the post-processing code
described in Refs. \cite{RBG91b,KWG94}. It was found in these
calculations that the nucleosynthesis yields for the weak $\it{s}$
process show large variations due to the uncertainties of the
involved neutron capture cross sections \cite{HKU08a}. Unlike in the
main $s$ process, flow equilibrium is not reached during the weak
$s$ process. Therefore, the neutron capture cross sections do not
only influence the yield of the respective isotope, but also the
production of all heavier nuclei on the reaction path of the weak
$s$-process. The impact of the improved MACS described here are
illustrated in Fig. \ref{fig03}.

The range of uncertainties, which are caused by the extrapolation
from the measured energy at $kT = 25$ keV to the higher energies
around $kT = 90$ keV are in some cases very large. To improve this
situation, complementary new accurate TOF measurements are clearly
needed in the mass region $56 \leq A \leq 70$. Accordingly, one has
to conclude that a reliable abundance predictions for the weak $s$
process are only possible if all neutron capture cross sections of
the involved isotopes are known with high accuracy. Also the
abundant light isotopes below Fe are important, since they may
constitute crucial neutron poisons for the $\it{s}$ process.

\section*{Acknowledgments} 
We are thankful to D. Roller, E.-P. Knaetsch, and W. Seith for their
support during the measurements. M.P and R.G. acknowledge support by
the Italian MIUR-PRIN06 Project "Late phases of Stellar Evolution:
Nucleosynthesis in Supernovae, AGB stars, Planetary Nebulae". M.P.
is supported by the Marie Curie Int. Reintegr. Grant
MIRG-CT-2006-046520 within the European FP6, and by NSF grants PHY
02-16783 (JINA).


\begin{thebibliography}{}

\bibitem{BBF57}
Burbidge, E.M., Burbidge, G.R., Fowler, W.A., \& Hoyle, F.
\newblock 1957, Rev. Mod. Phys., 29, 547

\bibitem{SGB95}
Straniero, O., Gallino, R., Busso, M., Chieffi, A., Raiteri, C.M.,
Limongi, M.,
  \& Salaris, M.
\newblock 1995, Ap. J., 440, L85

\bibitem{AKW99b}
Arlandini, C., K{\"a}ppeler, F., Wisshak, K., Gallino, R., Lugaro,
M., Busso,
  M., \& Straniero, O.
\newblock 1999, Ap. J., 525, 886 -- 900

\bibitem{RGB93}
Raiteri, C.M., Gallino, R., Busso, M., Neuberger, D., \&
K{\"a}ppeler, F.
\newblock 1993, Ap. J., 419, 207 -- 223

\bibitem{RHH02}
Rauscher, T., Heger, A., Hoffman, R.D., \& Woosley, S.E.
\newblock 2002, Ap. J., 576, 323

\bibitem{TEM07}
The, L.-S., El Eid, M., \& Meyer, B. 2007, Ap. J., 655, 1058

\bibitem{AnG89}
Anders, E. \& Grevesse, N.
\newblock 1989, Geochim. Cosmochim. Acta, 53, 197

\bibitem{SCL03}
Sneden, C., Cowan, J.C., Lawler, J.E., Ivans, I.I., Burles, S.,
Beers, T.C.,
  Primas, F., Hill, V., Truran, J.W., Fuller, G.M., Pfeiffer, B., \& Kratz,
  K.-L.
\newblock 2003, Ap. J., 591, 936 -- 953

\bibitem{TGA04}
Travaglio, C., Gallino, R., Arnone, E., Cowan, J., Jordan, F., \&
Sneden, C.
\newblock 2004, Ap. J., 601, 864

\bibitem{PGM08}
Pignatari, M., Gallino, R., Meynet, G., Hirschi, R., Herwig, F.,
Wiescher, M.,
\newblock 2008, Ap. J., 687, L95

\bibitem{FML06}
Fr{\"o}hlich, C., Martinez-Pinedo, G., Liebend{\"o}rfer, M.,
Thielemann, F.-K.,
  Bravo, E., Hix, W.R., Langanke, K., \& Zinner, N.T.
\newblock 2006, Phys. Rev. Lett., 96, 142502

\bibitem{Zin98}
Zinner, E.
\newblock 1998, Ann. Rev. Earth Planet. Sci., 26, 147 -- 188

\bibitem{NPA05}
Nassar, H., Paul, M., Ahmad, I., Berkovits, D., Bettan, M., Collon,
P.,
  Dababneh, S., Ghelberg, S., Greene, J.P., Heger, A., Heil, M., Henderson,
  D.J., Jiang, C.L., K{\"a}ppeler, F., Koivisto, H., O'Brien, S., Pardo, R.C.,
  Patronis, N., Pennington, T., Plag, R., Rehm, K.E., Reifarth, R., Scott, R.,
  Sinha, S., Tang, X., \& Vondrasek, R.
\newblock 2005, Phys. Rev. Lett., 94, 092504

\bibitem{BeK80}
Beer, H. \& K{\"a}ppeler, F.
\newblock 1980, Phys. Rev. C, 21, 534 -- 544

\bibitem{HKU08a}
Heil, M., K{\"a}ppeler, F., Uberseder, E., Gallino, R., \&
Pignatari, M.
\newblock 2008b, Phys. Rev. C, 77, 015808

\bibitem{SiC01}
Singh, B. \& Cameron, J.A.
\newblock 2001, Nucl. Data Sheets, 92, 1

\bibitem{Wu00}
Wu, S.-C.
\newblock 2000, Nuclear Data Sheets, 91, 1

\bibitem{Chu02}
Chunmei, Z.
\newblock 2002, Nucl. Data Sheets, 95, 59

\bibitem{Mac84b}
Macklin, R.L.
\newblock 1984, Nucl. Sci. Eng., 88, 129

\bibitem{KAM77}
Kenny, M., Allen, B.J., \& Macklin, R.L.
\newblock 1977, Aust. J. Phys., 30, 605

\bibitem{BBK00}
Bao, Z.Y., Beer, H., K{\"a}ppeler, F., Voss, F., Wisshak, K., \&
Rauscher, T.
\newblock 2000, Atomic Data Nucl. Data Tables, 76, 70 -- 154

\bibitem{UHK07}
Uberseder, E., Heil, M., K{\"a}ppeler, F., G{\"o}rres, J., \&
Wiescher, M.
\newblock 2007, Phys. Rev. C, 75, 0358019

\bibitem{HKU08b}
Heil, M., K{\"a}ppeler, F., Uberseder, E., Gallino, R., Bisterzo,
S., \&
  Pignatari, M.
\newblock 2008a, Phys. Rev. C, 78, 025802

\bibitem{MAM78b}
de~L.~Musgrove, A.R., Allen, B.J., \& Macklin, R.L.
\newblock 1978, in Neutron Physics and Nuclear Data for Reactors and other
  Applied Purposes,  (Paris: OECD) , 426

\bibitem{WBK86b}
Walter, G., Beer, H., K{\"a}ppeler, F., Reffo, G., \& Fabbri, F.
\newblock 1986, Astron. Astrophys., 167, 186 -- 199

\bibitem{Bur95}
Burrows, T.W.
\newblock 1995, Nucl. Data Sheets, 74, 1

\bibitem{Kat99}
Katakura, J.
\newblock 1999, Nucl. Data Sheets, 86, 955

\bibitem{Rei94}
Reich, C.W.
\newblock 1994, Nuclear Data Sheets, 71, 709

\bibitem{Hel98}
Helmer, R.G.
\newblock 1998, Nuclear Data Sheets, 83, 285

\bibitem{Hel94}
Helmer, R.G.
\newblock 1994, Nuclear Data Sheets, 72, 83

\bibitem{ReH00}
Reich, C.W. \& Helmer, R.G.
\newblock 2000, Nuclear Data Sheets, 90, 645

\bibitem{KWM85}
K{\"a}ppeler, F., Walter, G., \& Mathews, G.
\newblock 1985, Ap. J., 291, 319 -- 327

\bibitem{MSB99}
Mohr, P., Sedyshev, P.V., Beer, H., Stadler, W., Oberhummer, H.,
Popov, Yu.P.,
  \& Rochow, W.
\newblock 1999, Phys. Rev. C, 59, 3410

\bibitem{RBG91b}
Raiteri, C.M., Busso, M., Gallino, R., \& Picchio, G.
\newblock 1991, Ap. J., 371, 665 -- 672

\bibitem{KWG94}
K{\"a}ppeler, F., Wiescher, M., Giesen, U., G{\"o}rres, J., Baraffe,
I.,
  El~Eid, M., Raiteri, C.M., Busso, M., Gallino, R., Limongi, M., \& Chieffi,
  A.
\newblock 1994, Ap. J., 437, 396 -- 409
















\end{thebibliography}
\end{document}